\documentclass[sigconf, review, anonymous=false]{acmart}

\usepackage{xurl}
\usepackage[show]{chato-notes}
\usepackage[noadjust]{marginnote}
\hypersetup{
    colorlinks=true,
    linkcolor=blue,
    urlcolor=blue,
    breaklinks=true
}

\AtBeginDocument{%
  }

\newcommand{\rev}[1]{\textcolor{black}{#1}}
\newcommand{\ya}[1]{\textcolor{black}{#1}}
\newcommand{\io}[1]{\textcolor{black}{#1}}
\newcommand{\gm}[1]{\textcolor{black}{#1}}

\newcommand{\pageenlarge}[1]{\marginnote{}\enlargethispage{#1\baselineskip}}

\copyrightyear{2026}
\acmYear{2026}
\setcopyright{none}
\acmConference[WExIR @ SIGIR '26]{Workshop on Explainability in Information Retrieval at SIGIR 2026}{July 24, 2026}{Melbourne, Naarm, Australia}
\acmBooktitle{WExIR @ SIGIR '26, July 24, 2026, Melbourne, Naarm, Australia}
\acmDOI{}
\acmISBN{}

\begin{document}

\title{Explaining When PRF Fails: Participatory Auditing for Selective Query Expansion}

\author{Zeyan Liang}
\affiliation{%
  \institution{University of Glasgow}
  \city{Glasgow}
  \country{Scotland, UK}
}
\email{z.liang.1@research.gla.ac.uk}

\author{Graham McDonald}
\affiliation{%
  \institution{University of Glasgow}
  \city{Glasgow}
  \country{Scotland, UK}
}
\email{graham.mcdonald@glasgow.ac.uk}

\author{Iadh Ounis}
\affiliation{%
  \institution{University of Glasgow}
  \city{Glasgow}
  \country{Scotland, UK}
}
\email{iadh.ounis@glasgow.ac.uk}

\begin{CCSXML}
<ccs2012>
   <concept>
       <concept_id>10002951.10003317.10003338</concept_id>
       <concept_desc>Information systems~Retrieval models and ranking</concept_desc>
       <concept_significance>500</concept_significance>
   </concept>
</ccs2012>
\end{CCSXML}

\ccsdesc[500]{Information systems~Retrieval models and ranking}
\vspace{-3mm}
\keywords{PRF, Explainability, Participatory Auditing, Evaluation}

\begin{abstract}
\looseness -1 Pseudo-Relevance Feedback (PRF) improves retrieval effectiveness on average, but harms a substantial fraction of queries through \textit{query drift}, an asymmetry hidden by aggregate offline metrics.
Existing Selective PRF (sPRF) approaches typically rely on Query Performance Prediction (QPP) \rev{methods derived from the same ranking statistics, and therefore inherit, rather than resolve, this opacity}.
We argue that this is a core explainability problem in \io{IR}, and propose a two-stage \textit{audit-then-automate} framework.
In Stage~1, a participatory audit with 108 users across 43 TREC Deep Learning 2019 queries shows that only 20.9\% of queries benefit from PRF, while 25.6\% suffer \io{a} degraded user experience, and that avoiding harm is nearly twice as valuable as exploiting successful expansion.
In Stage~2, we repurpose LLM-based rerankers as \textit{system preference predictors} that replicate these user-derived labels automatically, grounded in inspectable document evidence. \rev{Together, the two stages explain \textit{which} queries PRF harms, \textit{why} an sPRF decision is made, and \textit{how} the decision can be inspected at scale, turning an opaque retrieval component into an auditable, user-grounded one.}
\vspace{-2mm}
\end{abstract}

\maketitle
\vspace{-2mm}
\thispagestyle{empty}
\section{Introduction}
\pageenlarge 1
Pseudo-Relevance Feedback (PRF) is a widely adopted technique for enhancing retrieval effectiveness by expanding queries with terms extracted from top-ranked documents~\cite{Rocchio1971, lavrenko2001relevance}.
Although PRF often results in average relevance gains, it suffers from a well-known robustness issue: \textit{query drift}, where non-relevant expansion terms shift the ranking away from the user's original intent~\cite{carpineto2012survey, zighelnic2008query}.
This drift creates an asymmetric per-query impact: while some queries benefit, others suffer from ranking degradation~\cite{turpin2006user}.
This per-query asymmetry is hidden at two levels.
For \textit{system designers}, aggregate offline metrics such as nDCG average performance over queries and conceal \ya{the} specific queries \ya{that are} degraded by \ya{query} expansion \ya{as well as the magnitude of such a degradation}~\cite{al2007relationship, turpin2006user}.
For \textit{end users}, the system gives no signal that expansion was applied at all, and no explanation for why their search results have changed.
We argue that this two-sided opacity is a core \textit{explainability} problem \ya{for automatic query expansion} in \ya{Information Retrieval (IR)}: making PRF accountable requires per-query explanations that are interpretable to both system designers (who decide whether to ship expansion) and \ya{end-users} (who must trust the results that are returned).

Selective PRF (sPRF) aims to mitigate query drift by applying expansion only to \gm{the} queries that are predicted to benefit \gm{in terms of retrieval} \ya{performance}, typically using Query Performance Prediction (QPP) \rev{methods}~\cite{shtok2012predicting}.
However, QPP methods lack robustness across diverse retrieval tasks~\cite{chifu2025uncovering} and are validated against the same offline \gm{metrics} that often fail to reflect \ya{actual} user satisfaction~\cite{turpin2006user}, providing no transparent justification for \textit{when} or \textit{why} expansion is applied to a given query.
The \ya{applied} selection mechanism, therefore, inherits, rather than resolves, the explainability gap.

\looseness -1 \io{We} propose a two-stage \textit{audit-then-automate} framework that closes this gap by producing explanations that are useful to \ya{system designers and end-users}.
In Stage~1, a participatory audit~\cite{liang2026auditing} transforms user interactions into transparent, query-level signals that reveal \textit{which} queries PRF harms and \textit{why}, providing system designers with auditable evidence of when expansion fails.
In Stage~2, an automated closed-loop predictor repurposes LLM-based document rerankers as system preference predictors that replicate these user-derived signals at deployment time, so that selective PRF decisions can be made for new queries without re-running user studies, while remaining traceable to \io{documents} that an \ya{end-user} can inspect.
\vspace{-2mm}
\section{Stage 1: Participatory Audit of PRF}
\label{sec:audit}
\begin{figure}[t]
  \centering
  \includegraphics[width=0.85\columnwidth]{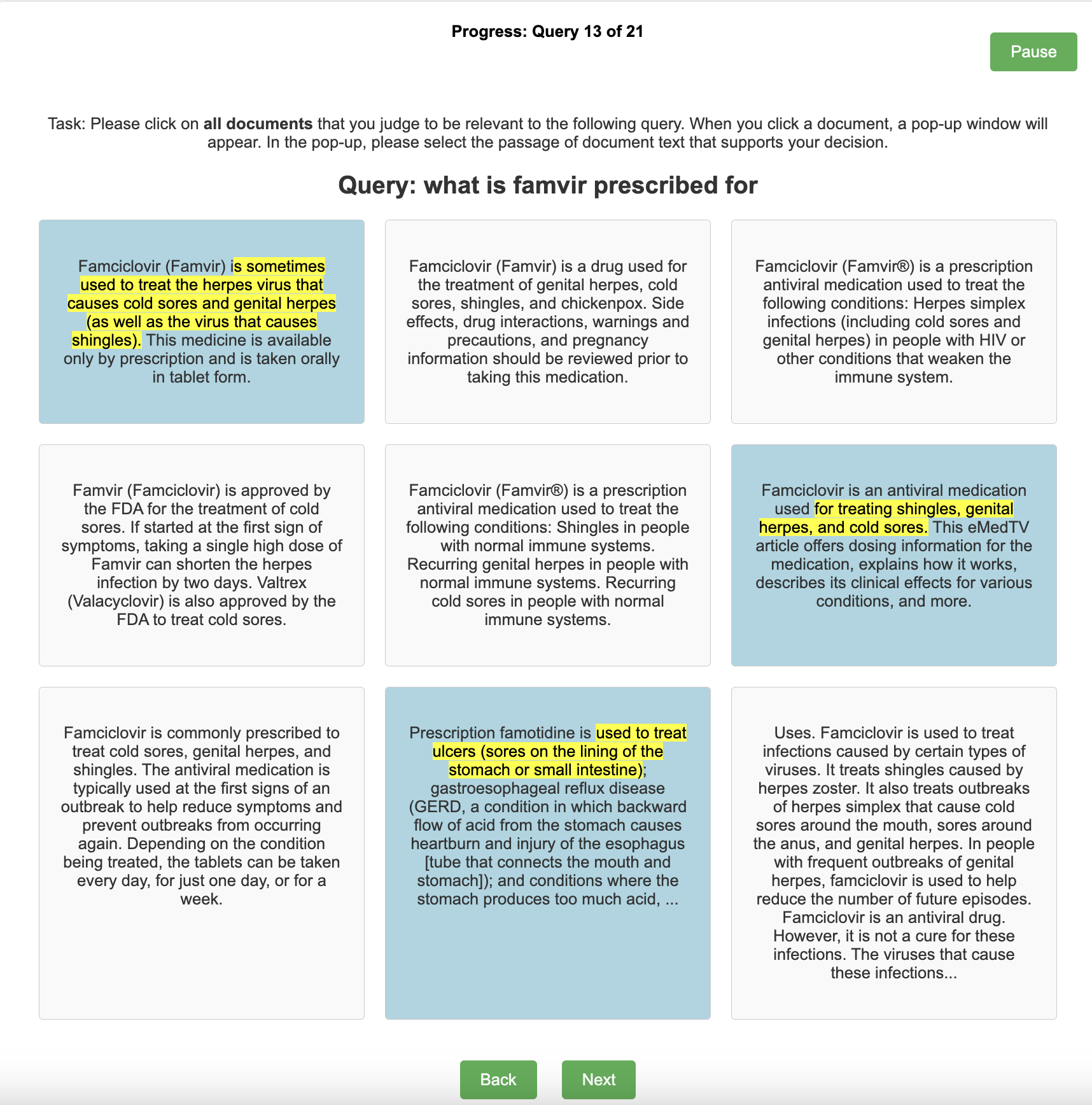}
  \vspace{-3mm}
  \caption{Our user interface for collecting audit signals.}
  \label{fig:interface}
  \vspace{-5mm}
\end{figure}

\pageenlarge 1
Our participatory audit is designed to make the per-query impact of PRF visible by collecting preference signals from users' natural search behaviour, without requiring users to complete an effortful relevance annotation task~\cite{liang2026auditing}.
For each of \ya{the} 43 TREC Deep Learning 2019 queries, we generate dual rankings \io{using} ColBERT~\cite{khattab2020colbert} (baseline) and ColBERT-PRF~\cite{wang2023colbert} (expanded), and construct a fair document set using Team Draft Interleaving (TDI)~\cite{radlinski2006minimally}, which treats the two rankings as `captains' in a sports draft and alternates selections to produce a combined set with an equal number of documents from each ranking. We present this combined document set in a 3$\times$3 grid \ya{using} Latin Square ordering~\cite{kammerer2014role, mackenzie2024human} to mitigate position bias.
As shown in Figure~\ref{fig:interface}, users perform a two-stage interaction: first clicking a document card to inspect its content in a pop-up, then selecting a specific passage to confirm relevance, \ya{allowing us to capture} high-confidence preference signals from natural browsing \io{behaviour}. \io{Each} document's originating system is recorded but hidden from users, such that user preference can be attributed to retrieval quality rather than rank position.

\looseness -1 We conduct two sequential user studies, \gm{each with 54 mutually exclusive participants}~\cite{liang2026auditing}. \rev{\gm{The first study} audits per-query PRF impact\gm{s} by collecting preferences over interleaved baseline/PRF rankings, identifying which queries \gm{have benefited}, or were harmed, \gm{by PRF}. \gm{The second study} then validates \gm{the effectiveness of} selective PRF.} Together, the \gm{user studies} \io{address} three explainability questions: \textit{which} queries are harmed by PRF, \rev{\textit{whether} selectively not applying PRF on harmed queries contributes more than applying it on beneficial ones}, and \textit{when} expansion should be applied.

% \pageenlarge 1
\looseness -1 First, only 20.9\% of queries benefit from PRF, while 25.6\% suffer degraded user experience, and the majority (53.5\%) show no clear preference; this heterogeneity, entirely hidden from aggregate offline evaluation, becomes visible and explainable at the per-query level through the audit.
Second, eliminating harmful expansions (i.e., reverting to the baseline ranking \gm{for queries that are degraded by PRF, i.e.,} PRF-Hurt queries) yields a 68.5\% relative improvement in user engagement, nearly twice the 37.0\% gain obtained by applying PRF only to the queries identified as beneficial (PRF-Benefit \gm{queries}),
%\inote{unclear what amplifying beneficial PRF means}
which provides a concrete, user-grounded explanation for why sPRF should be preferred: avoiding query drift is more valuable than exploiting successful expansion, consistent with prior evidence that users are disproportionately sensitive to query failures~\cite{turpin2006user}.
Third, applying PRF only to queries \io{for which} the audit signals indicate benefit achieves a significant gain of +10.5
%\inote{is that on average? be clear about what 10.5 means in this context}
adjusted selections per query, averaged across the 20 selective queries (11 PRF-Hurt + 9 PRF-Benefit) ($p$\,{=}\,0.032, Cohen's $d$\,{=}\,1.9) over universal PRF application. \io{This result is} validated through a Difference-in-Differences~\cite{angrist2009mostly} analysis with Neutral queries \io{as a control}
%\inote{explain what neutral queries mean}
(those queries whose PRF preference ratio falls within the \io{[0.45, 0.55]} tolerance band, \io{with} no clear user preference; \io{the} interleaved ranking is held constant across both user studies, so any change in engagement reflects cohort-level drift rather than the sPRF decision). Together, these results show that transforming users' natural interactions into structured audit signals produces per-query explanations of PRF impact that are both transparent to system designers (\textit{which} queries fail) and actionable for selective expansion (\textit{when} to apply PRF).
\vspace{-2mm}
\section{Stage 2: Closing the Loop with LLM Rerankers}
\label{sec:automate}

\looseness -1 The participatory audit produces ground-truth, user-derived explanations, but cannot be repeated for every new query. \ya{To} scale the approach, we argue that LLM-based document rerankers can serve as inspectable surrogates for user preference.
The key idea is to repurpose rerankers as \textit{system preference predictors}: rather than asking an LLM to classify whether PRF helps or \ya{harms} a query \gm{(an opaque process)}, we propose to present the LLM with the same interleaved document pool that users saw in Stage~1, and \ya{use} its ranking of these documents \ya{to} reveal which system it prefers.

%We argue that 

\looseness -1 \gm{This design enables explainability and transparency in selective PRF.} The LLM's output ranking is itself the explanation: an sPRF decision can be justified by pointing to specific documents that the LLM ranked highly from each system, mirroring how users in Stage~1 revealed preferences through their selections rather than through explicit judgements.
This supports inspection at two levels: for system designers, an auditable trace of which documents drove each sPRF decision; for \ya{end-users}, \ya{we propose} an in-interface rationale, for example by highlighting the supporting passages that distinguish the chosen system from its alternative, \ya{thereby} replacing an opaque expansion on/off switch with a grounded reason for the displayed results.
We see this as a path from a one-off audit study to an always-on transparency mechanism in deployed retrieval systems, and as a step towards selective PRF, whose decisions are explainable by construction rather than \ya{through} post-hoc rationalisation.

\vspace{-2.6mm}
\section{Discussion: Explainability Through Auditing}
\label{sec:discussion}

\looseness -1 Our framework contributes to explainability in IR in two complementary ways.
First, it makes per-query harms visible: standard offline evaluation averages over \ya{all} queries and conceals that over a quarter of queries are degraded by PRF, while our audit produces per-query labels that explicitly identify \textit{which} queries are harmed and \textit{to what extent} they are harmed, providing the granular transparency that regulatory frameworks such as the EU Digital Services Act increasingly require~\cite{EU20222065}.
Second, our \ya{generated} explanations are derived directly from user behaviour, i.e., users selecting more useful passages from one system's documents than from the other's, and the same evidence base, namely the interleaved documents themselves, is then re-used by the LLM reranker \ya{at deployment time} to replicate this judgement.
Every automated sPRF decision is therefore traceable end-to-end, from a query's interleaved documents to a user-grounded preference signal to an inspectable predictor output, \ya{constituting} a form of process-level explainability that is rare in current IR systems \ya{despite its importance}.
\vspace{-2.6mm}
\section{Conclusions}

\looseness -1 \ya{We} have presented a two-stage \textit{audit-then-automate} framework that transforms Pseudo-Relevance Feedback from an opaque retrieval component into one whose per-query decisions are transparent, user-grounded, and automatically reproducible. Our participatory audit with 108 users shows that avoiding harm is nearly twice as valuable as exploiting successful expansion on the queries that benefit from it, while our initial LLM-based automation shows that these user-derived preferences can be replicated at scale. By mapping each sPRF decision to those documents that users inspected in the audit, and that the reranker preferred at deployment, the framework provides explanations that are auditable to system designers and interpretable to \ya{end-users}. We see this \textit{audit-then-automate} paradigm as a step towards explainability in IR that does not stop at rationalising algorithmic outputs in isolation, but instead explains an algorithm's \textit{impact on real users}.

\balance
\bibliographystyle{ACM-Reference-Format}
\bibliography{wexir2026}

\end{document}